\documentclass[12pt]{article}
\usepackage[table]{xcolor}
\usepackage{colortbl}
\definecolor{lightgray}{gray}{0.9}
\usepackage{graphicx}
\usepackage{lscape}
\usepackage{citesort}
\usepackage{amssymb}
\usepackage{appendix}
\usepackage{multirow}

\usepackage{graphicx}
\usepackage{lscape}
\usepackage{citesort}
\usepackage{amssymb}
\usepackage{appendix}
\usepackage{multirow}

\usepackage{mathrsfs}
\begin{document}
\def\qq{\langle \bar q q \rangle}
\def\uu{\langle \bar u u \rangle}
\def\dd{\langle \bar d d \rangle}
\def\sp{\langle \bar s s \rangle}
\def\GG{\langle g_s^2 G^2 \rangle}
\def\Tr{\mbox{Tr}}
\def\figt#1#2#3{
        \begin{figure}
        $\left. \right.$
        \vspace*{-2cm}
        \begin{center}
        \includegraphics[width=10cm]{#1}
        \end{center}
        \vspace*{-0.2cm}
        \caption{#3}
        \label{#2}
        \end{figure}
    }

\def\figb#1#2#3{
        \begin{figure}
        $\left. \right.$
        \vspace*{-1cm}
        \begin{center}
        \includegraphics[width=10cm]{#1}
        \end{center}
        \vspace*{-0.2cm}
        \caption{#3}
        \label{#2}
        \end{figure}
                }

\def\ds{\displaystyle}
\def\beq{\begin{equation}}
\def\eeq{\end{equation}}
\def\bea{\begin{eqnarray}}
\def\eea{\end{eqnarray}}
\def\beeq{\begin{eqnarray}}
\def\eeeq{\end{eqnarray}}
\def\ve{\vert}
\def\vel{\left|}
\def\ver{\right|}
\def\nnb{\nonumber}
\def\ga{\left(}
\def\dr{\right)}
\def\aga{\left\{}
\def\adr{\right\}}
\def\lla{\left<}
\def\rra{\right>}
\def\rar{\rightarrow}
\def\lrar{\leftrightarrow}
\def\nnb{\nonumber}
\def\la{\langle}
\def\ra{\rangle}
\def\ba{\begin{array}}
\def\ea{\end{array}}
\def\tr{\mbox{Tr}}
\def\ssp{{\Sigma^{*+}}}
\def\sso{{\Sigma^{*0}}}
\def\ssm{{\Sigma^{*-}}}
\def\xis0{{\Xi^{*0}}}
\def\xism{{\Xi^{*-}}}
\def\qs{\la \bar s s \ra}
\def\qu{\la \bar u u \ra}
\def\qd{\la \bar d d \ra}
\def\qq{\la \bar q q \ra}
\def\gGgG{\la g^2 G^2 \ra}
\def\q{\gamma_5 \not\!q}
\def\x{\gamma_5 \not\!x}
\def\g5{\gamma_5}
\def\sb{S_Q^{cf}}
\def\sd{S_d^{be}}
\def\su{S_u^{ad}}
\def\sbp{{S}_Q^{'cf}}
\def\sdp{{S}_d^{'be}}
\def\sup{{S}_u^{'ad}}
\def\ssp{{S}_s^{'??}}

\def\sig{\sigma_{\mu \nu} \gamma_5 p^\mu q^\nu}
\def\fo{f_0(\frac{s_0}{M^2})}
\def\ffi{f_1(\frac{s_0}{M^2})}
\def\fii{f_2(\frac{s_0}{M^2})}
\def\O{{\cal O}}
\def\sl{{\Sigma^0 \Lambda}}
\def\es{\!\!\! &=& \!\!\!}
\def\ap{\!\!\! &\approx& \!\!\!}
\def\md{\!\!\!\! &\mid& \!\!\!\!}
\def\ar{&+& \!\!\!}
\def\ek{&-& \!\!\!}
\def\kek{\!\!\!&-& \!\!\!}
\def\cp{&\times& \!\!\!}
\def\se{\!\!\! &\simeq& \!\!\!}
\def\eqv{&\equiv& \!\!\!}
\def\kpm{&\pm& \!\!\!}
\def\kmp{&\mp& \!\!\!}
\def\mcdot{\!\cdot\!}
\def\erar{&\rightarrow&}
\def\olra{\stackrel{\leftrightarrow}}
\def\ola{\stackrel{\leftarrow}}
\def\ora{\stackrel{\rightarrow}}

\def\simlt{\stackrel{<}{{}_\sim}}
\def\simgt{\stackrel{>}{{}_\sim}}


\title{
         {\Large
                 {\bf
                     Analysis of the strong $D_{2}^{*}(2460)^{0}\rightarrow D^+
\pi^- $ and $D_{s2}^{*}(2573)^{+}\rightarrow D^{+} K^{0}$
transitions via  QCD sum rules
                 }
         }
      }

\author{\vspace{1cm}\\
{\small K. Azizi$^a$ \thanks {e-mail: kazizi@dogus.edu.tr}\,, Y.
Sarac$^b$
\thanks {e-mail: ysoymak@atilim.edu.tr}\,\,, H.
Sundu$^c$ \thanks {e-mail: hayriye.sundu@kocaeli.edu.tr}} \\
{\small $^a$  Physics Department, Do\u gu\c s University, Ac{\i}badem-Kad{\i}k\"oy, 34722 Istanbul, Turkey} \\
{\small $^b$ Electrical and Electronics Engineering Department,
Atilim University, 06836 Ankara, Turkey} \\
{\small $^c$ Department of Physics, Kocaeli University, 41380 Izmit,
Turkey}}
\date{}

\begin{titlepage}
\maketitle
\thispagestyle{empty}

\begin{abstract}
The strong $D_{2}^{*}(2460)^{0}\rightarrow
D^+ \pi^- $ and $D_{s2}^{*}(2573)^{+}\rightarrow D^{+} K^{0}$
transitions are analyzed via three point QCD sum rules. First, we
calculate  the corresponding strong  coupling constants
$g_{D_2^*D\pi}$ and $g_{D_{s2}^*DK}$. Then, we use them to
calculate  the corresponding decay widths and branching ratios. Making
use of the existing experimental data on the ratio of the decay
width in the pseudoscaler $D$ channel  to that of the vector $D^*$ channel,
finally, we estimate the decay width  and branching ratio of the strong $D_2^*(2460)^0
\rightarrow D^*(2010)^+\pi^-$ transition.

\end{abstract}

~~~PACS number(s): 11.55.Hx, 13.25.-k, 13.25.Ft
\end{titlepage}

\section{Introduction}
Following the first observation reported in 1986~\cite{Albrecht} the
past few decades have been a period for the observations of
orbitally excited charmed mesons
\cite{Albrecht1,Anjos,Frabetti,Avery,Alexander,Kubota,Bergfeld,Link,Abe,Abazov,Aaij}.
During this period there have also been several theoretical studies
on the masses, strong and electromagnetic transitions of these
mesons via various methods (for instance see \cite{Godfrey,Azizi,Azizi1,Azizi2} and  references therein).
  Among these orbitally excited mesons
are the $D_{2}^*(2460)$ and $D_{s2}^*(2573)$ mesons. The
$D_{2}^*(2460)$ state has the quantum numbers
$I(J^P)=\frac{1}{2}(2^+)$. Being not
known exactly, $I(J^P)=0(2^+)$ quantum numbers are favored by
the width and decay modes of the  $D_{s2}^*(2573)$ state. In this work, it is considered as a
charmed strange tensor meson. One can see
\cite{Fazio,Molina,Aubert1,Liventsev,Colangelo,Aubert2,Aubert3}
and references therein for some experimental and theoretical studies
on the properties of the charmed strange mesons.

In the literature, compared to the other types of mesons, there are
little theoretical works on the properties of the tensor mesons.
Especially, their strong transitions are not studied much. Studying
the parameters of these tensor mesons and the comparison of the
attained results with the existing experimental results may provide
fruitful information about the internal structures and the natures
of these mesons.
 Considering the appearance of these
charmed tensor mesons as  intermediate states in studying the $B$ meson decays,
the results of this work can also be helpful in this respect. Beside all of these, the possibility for
searches on the decay properties of $D_{2}^*$ and $D_{s2}^*$ mesons at LHC is another
motivation for theoretical studies on these states.

The present work puts forward the analysis of the strong  transitions
$D_{2}^{*}(2460)^{0}\rightarrow D^+ \pi^-$ and
$D_{s2}^{*}(2573)^{+}\rightarrow D^{+} K^0$. For this aim, first we calculate the strong coupling form factors $g_{D_2^*D\pi}$ and $g_{D_{s2}^*DK}$ via QCD sum rules as one of the most powerful and applicable non-perturbative
methods to hadron physics \cite{Shifman}.
These strong coupling form factors are then used to
calculate  the corresponding  decay widths and branching ratios of the transitions under consideration. Making
use of the existing experimental data on the ratio of the decay
width in the pseudoscaler $D$ channel  to that of the vector $D^*$ channel,
finally, we  evaluate the decay width  of the strong $D_2^*(2460)^0
\rightarrow D^*(2010)^+\pi^-$ transition.


\section{ QCD sum rules for the strong coupling form factors $g_{D_2^*D\pi}$ and $g_{D_{s2}^*DK}$ }
The aim of this section is to present the details of the
calculations of the coupling form factors $g_{D_2^*D\pi}$ and $g_{D_{s2}^*DK}$ for which we use the
following three-point correlation function:
\begin{eqnarray}\label{CorrelationFunc}
\Pi_{\mu\nu}(p,p^{\prime},q)=i^2 \int d^4x~ \int d^4y~e^{-ip\cdot x}~
e^{ip^{\prime}\cdot y}~{\langle}0| {\cal T}\left (
J^{D}(y)~J^{\pi[K]}(0)~
J^{D_{2}^{\ast^{\dag}}[D_{s2}^{\ast^{\dag}}]}_{\mu\nu}(x)\right)|0{\rangle},
\end{eqnarray}
were ${\cal T}$ is  the time ordering operator and $q=p-p'$ is
transferred momentum. The interpolating currents appearing in this
three-point correlation function can be written in terms of the
quark field operators as
\begin{eqnarray}\label{mesonintfield}
J^{D}(y)&=& i\bar{d}(y)\gamma_{5}c(y),
\nonumber \\
J^{\pi[K]}(0)&=& i\bar{u}[\bar{s}](0)\gamma_{5}d(0),
\nonumber \\
J_{\mu\nu}^{D_{2}^{\ast}[D_{s2}^{\ast}]}(x)&=&\frac{i}{2}\left[\bar
u[\bar s](x) \gamma_{\mu} \olra{\cal D}_{\nu}(x) c(x)+\bar u[\bar
s](x) \gamma_{\nu} \olra{\cal D}_{\mu}(x) c(x)\right],
\end{eqnarray}
with $ \olra{\cal D}_{\mu}(x)$ being the two-side covariant derivative
that acts on left and right, simultaneously. The covariant
derivative $ \olra{\cal D}_{\mu}(x)$  is defined as
\begin{eqnarray}\label{derivative}
\olra{\cal D}_{\mu}(x)=\frac{1}{2}\left[\ora{\cal D}_{\mu}(x)-
\ola{\cal D}_{\mu}(x)\right],
\end{eqnarray}
where
\begin{eqnarray}\label{derivative2}
\overrightarrow{{\cal
D}}_{\mu}(x)=\overrightarrow{\partial}_{\mu}(x)-i
\frac{g}{2}\lambda^aA^a_\mu(x),\nonumber\\
\overleftarrow{{\cal
D}}_{\mu}(x)=\overleftarrow{\partial}_{\mu}(x)+
i\frac{g}{2}\lambda^aA^a_\mu(x).
\end{eqnarray}
Here $\lambda^a$ ($a=1,~2,~ ....., ~8$) are the Gell-Mann
matrices and $A^a_\mu(x)$ stand for the external  gluon fields. These fields
are  expressed  in terms of the gluon field strength
tensor using the Fock-Schwinger gauge ( $x^\mu A^a_\mu(x)=0$), i.e.
\begin{eqnarray}\label{gluonfield}
A^{a}_{\mu}(x)=\int_{0}^{1}d\alpha \alpha x_{\beta}
G_{\beta\mu}^{a}(\alpha x)= \frac{1}{2}x_{\beta}
G_{\beta\mu}^{a}(0)+\frac{1}{3}x_\eta x_\beta {\cal D}_\eta
G_{\beta\mu}^{a}(0)+\cdots~,
\end{eqnarray}
where we keep only the leading term in our calculations and ignore from contributions of the derivatives of the gluon
field strength tensor.

One follows two different ways to calculate the above mentioned correlation
function according to the QCD sum rule approach. It is calculated in terms of hadronic parameters called hadronic side. On the other hand, it is calculated in terms of quark and gluon degrees of
freedom by the help of the operator product expansion  in deep
Euclidean region called the OPE side. The match of the coefficients of same structures
from both sides provides the QCD sum rules for the intended physical
quantities. By the help of double Borel transformation with respect to
the variables $p^2$ and $p'^2$  one suppresses the contribution of
the higher states and continuum.

In hadronic side,   the correlation function
in Eq.~(\ref{CorrelationFunc}) is saturated with  complete sets of
appropriate $D_{2}^{\ast}[D_{s2}^{\ast}]$, $\pi[K]$ and $D$ hadronic
states with the same quantum numbers as the used interpolating
currents. Performing the four-integrals
over $x$ and $y$ leads to
\begin{eqnarray} \label{physide}
\Pi_{\mu\nu}^{had}(p,p^{\prime},q)&=&\frac{\langle 0 \mid
 J^{\pi[K]}\mid \pi[K](q)\rangle \langle 0 \mid
 J^{D}\mid D(p^{\prime})\rangle \langle D_2^*[ D_{s2}^*](p,\epsilon) \mid
J_{\mu\nu}^{D_2^*[ D_{s2}^*]}\mid
0\rangle}{(p^2-m_{D_2^*[D_{s2}^*]}^2)(p^{\prime^2}-m_{D}^2)(q^2-m_{\pi[K]}^2)}
\nonumber \\
&\times& \langle \pi[K](q)D(p^{\prime})\mid
D_2^*[D_{s2}^*](p,\epsilon)\rangle+\cdots~,
\end{eqnarray}
where $\cdots$ represents the contributions of
the higher states and continuum. The matrix elements appearing in
this equation  are parameterized as follows:
\begin{eqnarray}\label{matriselement1}
\langle 0 \mid
 J^{\pi[K]}\mid \pi[K](q)\rangle=i\frac{m_{\pi[K]}^2 f_{\pi[K]}}{m_{d}+m_{u[s]}},
\end{eqnarray}
\begin{eqnarray}\label{matriselement2}
\langle 0 \mid
 J^{D}\mid D(p^{\prime})\rangle=i\frac{m_{D}^2 f_{D}}{m_{d}+m_{c}},
\end{eqnarray}
\begin{eqnarray}\label{matriselement3}
\langle D_2^*[ D_{s2}^*](p,\epsilon) \mid J_{\mu\nu}^{D_2^*}\mid
0\rangle=m_{D_2^*[ D_{s2}^*]}^3 f_{D_2^*[ D_{s2}^*]}
\epsilon_{\mu\nu}^{*(\lambda)}~,
\end{eqnarray}
and
\begin{eqnarray}\label{matriselement4}
\langle \pi[K](q)D(p^{\prime})\mid
D_2^*[D_{s2}^*](p,\epsilon)\rangle&=&g_{D_2^*D\pi[D_{s2}^*DK]}\epsilon^{(\lambda)}_{\eta\theta}~p^{\prime}_{\eta}~p^{\prime}_{\theta}~,
\end{eqnarray}
where $f_{\pi[K]}$, $f_{D}$ and $f_{D_2^*[D_{s2}^*]}$ are leptonic
decay constants of $\pi[K]$, $D$ and $D_2^*[D_{s2}^*]$ mesons,
respectively and $g_{D_2^*D\pi}$ and $g_{D_{s2}^*DK}$ are the strong coupling form factors among the mesons under consideration.
In writting  Eq. (\ref{matriselement4}) we have used  the following relationships of the polarization tensor
$\epsilon_{\eta\theta}^{(\lambda)}$ \cite{H.Y.Cheng}:
\begin{eqnarray}\label{polarizationtensorrelation}
\epsilon_{\eta\theta}^{(\lambda)}=\epsilon_{\theta\eta}^{(\lambda)},\,\,\,\,\,\epsilon_{\eta}^{(\lambda)\eta}=0,\,\,\,\,\,
p_{\eta}\epsilon^{\eta\theta}_{\lambda}=p_{\theta}\epsilon^{\eta\theta}_{\lambda}=0,\,\,\,\,\,
\epsilon_{\eta\theta}^{(\lambda)}\epsilon^{*(\lambda^{\prime})\eta\theta}=\delta_{\lambda\lambda^{\prime}}.
\end{eqnarray}
 By the usage of the matrix elements given in
Eqs.~(\ref{matriselement1}), (\ref{matriselement2}),
(\ref{matriselement3}) and (\ref{matriselement4}) in
Eq.~(\ref{physide}), the correlation function takes its final form
in the hadronic side,
\begin{eqnarray} \label{physide1}
\Pi_{\mu\nu}^{had}(p,p^{\prime},q)&=&\frac{g_{D_2^*D\pi[D_{s2}^*DK]}~m_D^2~
m_{\pi[K]}^2~f_D~f_{\pi[K]}~f_{D_2^*[D_{s2}^*]}}{(m_c+m_d)(m_{u[s]}+m_d)(p^2-m_{D_2^*[D_{s2}^*]}^2)
(p^{\prime^2}-m_D^2)(q^2-m_{\pi[K]}^2)}
\nonumber \\
&\times&\Big[ m_{D_2^*[D_{s2}^*]}p\cdot
p^{\prime}p_{\mu}^{\prime}~p_{\nu}-\frac{2~(p\cdot
p^{\prime})^2+m_{D_2^*[D_{s2}^*]}^2~p^{\prime^2}}{3~m_{D_2^*[D_{s2}^*]}}p_{\mu}~p_{\nu}-m_{D_2^*[D_{s2}^*]}^3p_{\mu}^{\prime}~p_{\nu}^{\prime}
\nonumber \\
&+& m_{D_2^*[D_{s2}^*]}(p\cdot p^{\prime})~p_{\mu}~p_{\nu}^{\prime}+
\frac{m_{D_2^*[D_{s2}^*]}(m_{D_2^*[D_{s2}^*]}^2~~p^{\prime^2}-(p\cdot
p^{\prime})^2)}{3}~g_{\mu\nu}\Big]+\cdots~,
\nonumber \\
\end{eqnarray}
where the  summation over the polarization
tensor has been applied, i.e.
\begin{eqnarray}\label{polarizationt1}
\sum_{\lambda}\varepsilon_{\mu\nu}^{(\lambda)}\varepsilon_{\alpha\beta}^{*(\lambda)}=\frac{1}{2}T_{\mu\alpha}T_{\nu\beta}+
\frac{1}{2}T_{\mu\beta}T_{\nu\alpha}
-\frac{1}{3}T_{\mu\nu}T_{\alpha\beta},
\end{eqnarray}
and
\begin{eqnarray}\label{polarizationt2}
T_{\mu\nu}=-g_{\mu\nu}+\frac{p_\mu p_\nu}{m_{D_2^*[D_{s2}^*]}^2}.
\end{eqnarray}

Following the application of the double Borel transformation with
respect to the initial and final momenta squared, we attain the
hadronic side of the correlation function as
\begin{eqnarray} \label{Borelphyside1}
\widehat{\textbf{B}}\Pi_{\mu\nu}^{had}(q)&=&g_{D_2^*D\pi[D_{s2}^*DK]}\frac{f_Df_{D_2^*[D_{s2}^*]}f_{\pi[K]}m_D^2
m_{\pi[K]}^2}{(m_c+m_d)(m_{u[s]}+m_d)(m_{\pi[K]}^2-q^2)}e^{-\frac{m_{D_2^*[D_{s2}^*]}^2}{M^2}}
e^{-\frac{m_D^2}{M^{\prime^2}}}
\nonumber \\
&\Bigg\{&\frac{1}{12}m_{D_2^*[D_{s2}^*]}
\Big(m_D^4+(m_{D_2^*[D_{s2}^*]}^2-q^2)^2-2m_D^2(m_{D_2^*[D_{s2}^*]}^2+q^2)\Big)g_{\mu\nu}
\nonumber \\
&+&\frac{1}{6m_{D_2^*[D_{s2}^*]}}\Big[m_D^4+m_D^2(4m_{D_2^*[D_{s2}^*]}^2-2q^2)
+(m_{D_2^*[D_{s2}^*]}^2-q^2)^2\Big]p_{\mu}p_{\nu}
\nonumber \\
&-&\frac{1}{2}m_{D_2^*[D_{s2}^*]}
(m_D^2+m_{D_2^*[D_{s2}^*]}^2-q^2)p_{\nu}p^{\prime}_{\mu}
+m_{D_2^*[D_{s2}^*]}^3p^{\prime}_{\mu}p^{\prime}_{\nu}
\nonumber \\
&-&\frac{1}{2}m_{D_2^*[D_{s2}^*]}
(m_D^2+m_{D_2^*[D_{s2}^*]}^2-q^2)p_{\mu}p^{\prime}_{\nu}
\Bigg\}+\cdots~,
\end{eqnarray}
where $M^2$ and $M^{\prime^2}$ are Borel mass parameters.

In  OPE side, we
calculate  the aforesaid correlation function  in
deep Euclidean region, where $p^2\rightarrow -\infty$ and $p'^2\rightarrow
-\infty$. Substituting
  the explicit forms of the interpolating currents into the
correlation function Eq.~(\ref{CorrelationFunc}) and after contracting out  all
quark pairs via Wick's theorem, we get
\begin{eqnarray}\label{correl.func.2}
\Pi^{OPE} _{\mu\nu}(p, p^{\prime}, q)&=&\frac{i^5}{2}\int d^{4}x\int
d^{4}ye^{-ip\cdot x}e^{ip^{\prime}\cdot y}
\nonumber \\
&\times& \Bigg\{Tr\left[\gamma_5~S_d^{ji}(-y)\gamma_5
S_c^{i\ell}(y-x)\gamma_\mu \olra{\cal D}_{\nu}(x) S_{u[s]}^{\ell
j}(x)\right]+ \left[\mu\leftrightarrow\nu\right] \Bigg\}~,
\end{eqnarray}
where $ S^{i\ell}_c(x)$ represents the heavy quark propagator which is given
by \cite{Reinders}
\begin{eqnarray}
S_{c}^{i\ell}(x)&=&\frac{i}{(2\pi)^4}\int d^4k e^{-ik \cdot x}
\left\{ \frac{\delta_{i\ell}}{\!\not\!{k}-m_c}
-\frac{g_sG^{\alpha\beta}_{i\ell}}{4}\frac{\sigma_{\alpha\beta}(\!\not\!{k}+m_c)+
(\!\not\!{k}+m_c)\sigma_{\alpha\beta}}{(k^2-m_c^2)^2}\right.\nonumber\\
&&\left.+\frac{\pi^2}{3} \langle \frac{\alpha_sGG}{\pi}\rangle
\delta_{i\ell}m_c \frac{k^2+m_c\!\not\!{k}}{(k^2-m_c^2)^4}
+\cdots\right\} \, ,
\end{eqnarray}
and $S_{u[s]}(x)$ and $S_d(x)$ are the light quark propagators and
are given by
\begin{eqnarray}\label{lightpropagator}
S_{q}^{ij}(x)&=& i\frac{\!\not\!{x}}{ 2\pi^2 x^4}\delta_{ij}
-\frac{m_q}{4\pi^2x^2}\delta_{ij}-\frac{\langle
\bar{q}q\rangle}{12}\Big(1 -i\frac{m_q}{4}
\!\not\!{x}\Big)\delta_{ij} -\frac{x^2}{192}m_0^2\langle
\bar{q}q\rangle\Big(1-i\frac{m_q}{6} \!\not\!{x}\Big)\delta_{ij}
\nonumber \\
&-&\frac{ig_s
G_{\theta\eta}^{ij}}{32\pi^2x^2}\big[\!\not\!{x}\sigma^{\theta\eta}+\sigma^{\theta\eta}\!\not\!{x}\big]
+\cdots \, .
\end{eqnarray}

After insertion of the explicit forms of the heavy and light
quark propagators into  Eq.~(\ref{correl.func.2}), we use the following transformations in $D=4$ dimensions:
\begin{eqnarray}\label{intyx}
\frac{1}{[(y-x)^2]^n}&=&\int\frac{d^Dt}{(2\pi)^D}e^{-it(y-x)}~i~(-1)^{n+1}~2^{D-2n}~\pi^{D/2}~
\frac{\Gamma(D/2-n)}{\Gamma(n)}\Big(-\frac{1}{t^2}\Big)^{D/2-n},
\nonumber \\
\frac{1}{[y^2]^m}&=&\int\frac{d^Dt'}{(2\pi)^D}e^{-it'y}~i~(-1)^{m+1}~2^{D-2m}~\pi^{D/2}~
\frac{\Gamma(D/2-m)}{\Gamma(m)}\Big(-\frac{1}{t^{'2}}\Big)^{D/2-m}
\end{eqnarray}
and  perform the
four-$x$ and four-$y$ integrals after the replacements $x_{\mu}\rightarrow
i\frac{\partial}{\partial p_{\mu}}$ and
 $y_{\mu}\rightarrow -i\frac{\partial}{\partial p'_{\mu}}$.
  The four-integrals over $k$ and $t'$ are performed by the help of the Dirac Delta functions which are obtained from the four-integrals over $x$ and $y$.
   The remaining four-integral over $t$ is performed via the Feynman parametrization and
\begin{eqnarray}\label{Int}
\int d^4t\frac{(t^2)^{\beta}}{(t^2+L)^{\alpha}}=\frac{i \pi^2
(-1)^{\beta-\alpha}\Gamma(\beta+2)\Gamma(\alpha-\beta-2)}{\Gamma(2)
\Gamma(\alpha)[-L]^{\alpha-\beta-2}}.
\end{eqnarray}
 Albeit its smallness we also include the contributions coming from
the two-gluon condensate in our calculations.

 The correlation function in OPE side is written in terms of different structures as
 \begin{eqnarray}\label{QCDside}
\Pi^{OPE}_{\mu\nu} (p, p^{\prime}, q)&=&\Pi_1(q^2)p_{\mu}p_{\nu}+
\Pi_2(q^2)p_{\nu}p^{\prime}_{\mu}+\Pi_3(q^2)p_{\mu}p^{\prime}_{\nu}+
\Pi_4(q^2)p^{\prime}_{\mu}p^{\prime}_{\nu}+\Pi_5(q^2)g_{\mu\nu}, \nonumber\\
\end{eqnarray}
where each $\Pi_i(q^2)$ function receives contributions
from both the perturbative and non-perturbative parts and can be written as
\begin{eqnarray}\label{QCDside1}
\Pi_i(q^2)=\int^{}_{}ds\int^{}_{}ds^{\prime}
\frac{\rho_i^{pert}(s,s^{\prime},q^2)}{(s-p^2)(s^{\prime}-p^{\prime^2})}+\Pi_i^{non-pert}(q^2),
\end{eqnarray}
where  the spectral densities $\rho_i(s,s',q^2)$ are
given by the imaginary parts of the $\Pi_{i}$ functions,
i.e., $\rho_i(s,s',q^2)=\frac{1}{\pi}Im[\Pi_{i}]$. In present
study, we consider the Dirac structure $p_{\mu}p_{\nu}$  to obtain the
QCD sum rules for the considered strong  coupling form factors. The $\rho_1(s,s',q^2)$ and
$\Pi_1^{non-pert}(q^2)$ corresponding to this Dirac structure are
obtained as
\begin{eqnarray}\label{rho}
\rho_1^{pert}(s,s^{\prime},q^2)&=& \int_{0}^{1}dx
\int_{0}^{1-x}dy\frac{3(1+8x^2-7y+8y^2-7x+16xy)}{8\pi^2}\theta[L(s,s^{\prime},q^2)],
\end{eqnarray}
with $\theta[...]$ being the unit-step function
and
\begin{eqnarray}\label{Nonpert1}
\Pi_1^{non-pert}(q^2)&=&\int_{0}^{1}dx \int_{0}^{1-x}dy
\Bigg\{\langle\frac{\alpha_{s}G^2}{\pi}\rangle\Bigg[\frac{1}{8L^4}m_cx^3(1-2x-2y)\Big[m_cm_dm_q(1-x-y)
\nonumber \\
&+&m_c
\Big(p^2x+q^2(y-1)\Big)(x+y-1)(x+y)+m_cp^{\prime^2}x(x+y-xy-y^2-1)
\nonumber \\
&+&\Big(m_q(x+y-1)-m_d(x+y)\Big)\Big(
p^2(x-1)(x+y-1)+y(p^{\prime^2}(1-x))
\nonumber \\
&+&q^2(x+y-1)\Big)
\Big]+\frac{1}{24L^3}\Big[(x-1)^2x^2(2x-1)\Big(p^{\prime^2}-q^2+p^2(3x-2)\Big)
\nonumber \\
&+&xy(x-1)\Big(q^2(x-1)
(4-13x+6x^2)+p^2(x-1)(2-17x+24x^2)
\nonumber \\
&+&p^{\prime^2}(3-11x+15x^2-6x^3)\Big)+q^2y^2(3-32x+81x^2-75x^3+24x^4)
\nonumber \\
&+&xy^2\Big(p^2(57x-90x^2+42x^3-10)+p^{\prime^2}(11-40x+50x^2-18x^3)\Big)+q^2y^3
\nonumber \\
&\times&(x-1)(15-62x+42x^2)+xy^3
\Big(p^2(x-1)(42x-19)+48xp^{\prime^2}-24x^2p^{\prime^2}
\nonumber \\
&-&19p^{\prime^2}\Big)+xy^4\Big(p^{\prime^2}(17-18x)-p^2(17-24x)\Big)
+q^2y^4(27-73x+42x^2)
\nonumber \\
&+&6xy^5(p^2-p^{\prime^2})+3y^5q^2(8x-7)+6y^6q^2-m_c^2x^3(1+8x^2-7y+8y^2-7x
\nonumber \\
&+& 16xy)-m_cm_qx(x+y-1)
(8x^3-3x^2-2x-5y+10xy+8x^2y+8y^2)
\nonumber \\
&+&m_cm_qx(8x^4-11x^3+8x^2-3x-3y+14xy-19x^2y+16x^3y+7y^2-12xy^2
\nonumber \\
&+&8x^2y^2-4y^3)
\Big]+\frac{1}{48L^2}\Big[24x^4+x^3(72y-55)+3x^2(13-48y+32y^2)
\nonumber \\
&+&(y^2-y)(8-31y+24y^2)-8x+75xy -144xy^2+72xy^3\Big]\Bigg]
\nonumber \\
&+&\frac{m_0^2\langle\overline{d}d\rangle
m_q}{24q^2(m_c^2-p^{\prime^2})^4}\Big(9m_c^4-8m_c^3m_d-12m_c^2p^{\prime^2}+2m_cm_dp^{\prime^2}
+3p^{\prime^4}\Big)
\nonumber \\
&+&\frac{m_0^2\langle\overline{q}q\rangle
m_d}{24q^2(m_c^2-p^2)^4}\Big(9m_c^4+8m_c^3m_q-12m_c^2p^{\prime^2}-2m_cm_qp^2
+3p^4\Big)
 \Bigg\},
\end{eqnarray}
where  $\langle\overline{q}q\rangle=\langle\overline{u}u\rangle$, $m_q=m_u$ and $\langle\overline{q}q\rangle=\langle\overline{s}s\rangle$, $m_q=m_s$ for the initial $D_2^*$ and $D_{s2}^*$ states, respectively and
\begin{eqnarray}\label{L}
L(s,s^{\prime},q^2)&=&
-m_c^2x+sx-sx^2+q^2y-q^2xy-sxy+s^{\prime}xy-q^2y^2.
\end{eqnarray}

The final form of the OPE side of the correlation function is
obtained after double Borel transformation as
\begin{eqnarray}\label{QCDsideBorel}
\widehat{\textbf{B}}\Pi^{OPE}_{\mu\nu}(q^2)=\Big\{\int^{}_{}ds\int^{}_{}ds^{\prime}e^{-\frac{s}{M^2}}
e^{-\frac{s^{\prime}}{M^{\prime^2}}}
\rho_1^{pert}(s,s^{\prime},q^2)+\widehat{\textbf{B}}\Pi_1^{nonpert}(q^2)\Big\}
p_{\mu}p_{\nu}+\cdots~,
\end{eqnarray}
where
\begin{eqnarray}\label{BorelQCD}
\widehat{\textbf{B}}\Pi^{non-pert}_{1}(q^2)&=&\int^{0}_{1}dx
\exp\Big[\frac{m_c^2M^{\prime^4}x+m_c^2M^4x+M^2M^{\prime^2}(-q^2(x-1)^2+2m_c^2x)}
{M^2M^{\prime^2}(M^2+M^{\prime^2})x(x-1)}\Big]\langle\frac{\alpha_{s}G^2}{\pi}\rangle
\nonumber \\
&\times& \frac{1}{48}\sqrt{\frac{1}{(x-1)^2}}\Bigg\{
\frac{M^{\prime^{12}}(x-1)^6
(M^2+M^{\prime^2}x)}{x^3u^6(M^2+M^{\prime^2})^{10}}\Big[xm_c^2(M^{\prime^4}+M^4)-M^{\prime^2}M^2
\nonumber \\
&\times& \Big(q^2(x-1)^2-2m_c^2x\Big)\Big]
+\frac{M^{\prime^{12}}(x-1)^6(M^2+M^{\prime^2}x)}{x^3u^5
(M^2+M^{\prime^2})^9}\Big(M^2q^2(x-1)
\nonumber \\
&+&4M^4x+M^{\prime^2}(q^2+2M^2x-q^2x)\Big)+\frac{M^{\prime^8}(x-1)^4}{x^2u^4M^2(M^2+M^{\prime^2})^7}
\Big[m_cm_dM^6
\nonumber \\
&+&
M^{\prime^6}x\Big(M^2(x-1)+m_cm_dx\Big)+M^4M^{\prime^2}\Big(4M^2(1-x)+m_cm_d(1+2x)
\Big)
\nonumber \\
&+&
M^2M^{\prime^4}\Big(m_cm_dx(2+x)+M^2(7x-5x^2-2)\Big)\Big]-\frac{M^{\prime^8}
(M^2+M^{\prime^2}x)}{x^2u^3M^2(M^2+M^{\prime^2})^5}
\nonumber \\
&\times&(x-1)^4
\Big[m_cm_u+M^2 \Big]
\Bigg\}\theta\Big[\frac{M^2-M^2x}{M^{\prime^2}+M^2}\Big]
\end{eqnarray}
with
\begin{eqnarray}\label{u}
u=-1+x+\frac{M^2-M^2x}{M^2+M^{\prime^2}}.
\end{eqnarray}
%

 Equating the coefficients of the same Dirac structure  from both sides of the correlation function,
 we get the following sum rules for the coupling form factors
$g_{D_2^*D\pi}$ and $g_{D_{s2}^*DK}$:
%
%
%
\begin{eqnarray}\label{couplingconstant}
g_{D_2^*D\pi[D_{s2}^*DK]}&=&e^{\frac{m_{D_2^*[D_{s2}^*]}^2}{M^2}}e^{\frac{m_D^2}{M^{\prime^2}}}~
\frac{6(m_c+m_d)(m_d+m_{u[s]})(m_{\pi[K]}^2-q^2)m_{D_2^*[D_{s2}^*]}}{f_{D_2^*[D_{s2}^*]}f_Df_{\pi[K]}
m_D^2m_{\pi[K]}^2}
\nonumber \\
&\times&\frac{1}{\Big[m_D^4+m_D^2(4m_{D_2^*[D_{s2}^*]}^2-2q^2)+(m_{D_2^*[D_{s2}^*]}^2-q^2)^2\Big]}~
\nonumber \\
&\times& \Bigg\{\int^{s_0}_{(m_c+m_{u[s]})^2}ds\int^{s'_0}_{(m_c+m_{d})^2}ds^{\prime}e^{-\frac{s}{M^2}}
e^{-\frac{s^{\prime}}{M^{\prime^2}}}
\rho_1^{pert}(s,s^{\prime},q^2)+\widehat{\textbf{B}}\Pi_1^{non-pert}(q^2)\Bigg\}~,\nonumber\\
\end{eqnarray}
where $s_0$ and $s'_0$ are continuum thresholds in $D_2^*[D_{s2}^*]$ and $D$ channels, respectively and we have used the quark-hadron duality assumption.

\section{Numerical Results}
In this part, we numerically analyze the obtained sum rules for the strong coupling form factors in the previous section and search for the behavior of those couplings
with respect to  $Q^2=-q^2$. The values of the strong coupling  form factors at $Q^2=-m_{\pi[K]}^2$ give the strong coupling constants whose values are then used to find the decay rate and branching
ratio of the strong transitions under consideration.  To go further, we use some input parameters presented in  Table 1.
\begin{table}[ht]\label{Table1}
\centering \rowcolors{1}{lightgray}{white}
\begin{tabular}{cc}
\hline \hline
   Parameters  &  Values
           \\
\hline \hline
$m_{c}$              & $(1.275\pm0.025)~\mbox{GeV}$\cite{Beringer}\\
$m_{d}$              & $4.8^{+0.5}_{-0.3}~\mbox{MeV}$\cite{Beringer}\\
$ m_{u} $            &$2.3^{+0.7}_{-0.5}~\mbox{MeV}$ \cite{Beringer}\\
$ m_{s} $            &   $ 95\pm5~\mbox{MeV}$ \cite{Beringer}\\
$ m_{D_2^*(2460)}$    &   $ (2462.6\pm0.6)~\mbox{MeV}$ \cite{Beringer}  \\
$ m_{D_{s2}^*(2573)} $      &   $ (2571.9\pm0.8)~\mbox{MeV}$  \cite{Beringer} \\
$ m_{D} $      &   $(1869.62\pm 0.15) ~\mbox{MeV} $ \cite{Beringer}  \\
$ m_{\pi} $      &   $(139.57018\pm 0.00035)~\mbox{MeV} $  \cite{Beringer} \\
$ m_{K} $      &   $(493.677\pm 0.016) ~\mbox{MeV} $  \cite{Beringer} \\
$ f_{D_2^*(2460)} $      &   $0.0228\pm0.0068 $  \cite{Azizi} \\
$ f_{D_{s2}^*(2573)} $      &   $0.023\pm0.0011 $  \cite{Azizi1} \\
$ f_{D} $      &   $206.7\pm8.9$ $\mbox{MeV}$  \cite{Beringer} \\
$ f_{\pi} $      &   $130.41\pm0.03\pm0.20 $ $\mbox{MeV}$ \cite{Beringer} \\
$ f_{K} $      &   $156.1\pm0.2\pm0.8\pm0.2 $  $\mbox{MeV}$ \cite{Beringer} \\
$ \langle\frac{\alpha_sG^2}{\pi}\rangle $       &   $(0.012\pm0.004)$ $~\mbox{GeV$^4$}$\cite{belyaev}   \\
 \hline \hline
\end{tabular}
\caption{Input parameters used in  calculations.}
\end{table}

The next task is to find the working regions for the auxiliary parameters $M^2$, $M'^2$, $s_0$ and
$s'_0$. Being not physical parameters, the strong coupling
form factors should roughly be independent of  these
parameters. In the case of the continuum thresholds, they are not completely arbitrary but are related to the energy of the first excited states with the same
quantum numbers as the considered interpolating fields. From numerical analysis, the working intervals are obtained as
 $7.6[8.5]~\mbox{GeV$^2$}\leq
s_0\leq8.8[9.4]~\mbox{GeV$^2$}$ and $4.7~\mbox{GeV$^2$}\leq
s'_0\leq5.6~\mbox{GeV$^2$}$ for the strong vertex $D_2^*D\pi[D_{s2}^*DK]$.
In the case of Borel mass parameters $M^2$ and $M'^2$, we choose their working windows such that they guarantee   not only the pole dominance but also the convergence of the OPE.  If  these parameters are chosen
 too large, the convergence of the
OPE is good but the continuum and higher state contributions
exceed the pole  contribution. On the other hand if one chooses
too small values, although the pole dominates  the higher state
and continuum contributions, the OPE have a poor convergence. By
considering these conditions we choose the windows
$3~\mbox{GeV$^2$}\leq M^2\leq 8~\mbox{GeV$^2$}$ and
$2~\mbox{GeV$^2$}\leq M'^2\leq 5~\mbox{GeV$^2$}$ for the Borel
mass parameters. Our analysis shows that, in these intervals, the
dependence of the results on the Borel parameters are weak.


Now we proceed to find the variations of the strong coupling form factors with respect to $Q^2$.  Using the working regions for the auxiliary parameters we observe that
 the following fit function well describes the strong coupling form factors in terms of $Q^2$:
\begin{eqnarray}\label{fitfunc}
g_{D_2^*D\pi[D_{s2}^*DK]}(Q^2)=c_1\exp\Big[-\frac{Q^2}{c_2}\Big]+c_3,
\end{eqnarray}
where the values of the parameters $c_1$, $c_2$ and $c_3$ for different structures are
presented in  tables \ref{fitparam} and \ref{fitparam1} for
$D_2^*D\pi$ and $D_{s2}^*DK$, respectively. From this fit
parametrization we obtain the values of the strong coupling
constants for each structure at $Q^2=-m_{\pi[K]}^2$ as
presented in table \ref{couplingconstant}. 
The errors appearing in
our results belong to the uncertainties in the input parameters as
well as errors coming from determination of the working regions for the
auxiliary parameters. 
From  table \ref{couplingconstant} we see that the results strongly depend on the selected structure such that the maximum values for the strong couplings in $D_2^*$ and $D_{s2}^*$ channels
 that belong to the structure
 $p^{\prime}_{\mu}p^{\prime}_{\nu}$ are roughly four times greater that those of the minimum values which correspond to the structure $p_{\mu}p_{\nu}$. The values obtained using other structures lie between these maximum and minimum values.
 Note that the coupling constant in $\pi$ channel has been  estimated  in a  pioneering
study via  Chiral perturbation theory \cite{Falk}. By converting  the parametrization of coupling constant used in \cite{Falk} to our parametrization,  \cite{Falk} finds a value of
 $g_{D_2^*D\pi}\simeq 16~GeV^{-1}$ in $\pi$ vertex which is close to  our  prediction obtained via the structure $g_{\mu\nu}$. Our results obtained via the structures $p^{\prime}_{\mu}p_{\nu}$
and $p_{\mu}p^{\prime}_{\nu}$
 are comparable with that of \cite{Falk}
 within the errors. However, our result obtained via the structure $p^{\prime}_{\mu}p^{\prime}_{\nu}$ are considerably high and our prediction obtained using the structure 
$p_{\mu}p_{\nu}$ is very low compared to the result of \cite{Falk} for the strong coupling constant associated to the $D_2^*D\pi$ vertex.

\begin{table}[h]
\renewcommand{\arraystretch}{1.5}
\addtolength{\arraycolsep}{3pt}
$$
\begin{array}{|c|c|c|c|}
\hline \hline
       \mbox{structure}    & c_1 (\mbox{GeV$^{-1}$})& c_2 (\mbox{GeV$^2$})&c_3 (\mbox{GeV$^{-1}$})    \\
\hline
  \mbox{$p_{\mu}p_{\nu}$} &5.17\pm 1.50&13.21\pm3.84&-(0.54\pm0.16) \\
  \hline
  \mbox{$p^{\prime}_{\mu}p^{\prime}_{\nu}$} &8.12\pm 2.34&11.14\pm2.78&12.56\pm3.77 \\
  \hline
   \mbox{$p^{\prime}_{\mu}p_{\nu}$} &11.57\pm 3.12&12.55\pm3.51&1.13\pm0.34 \\
  \hline
  \mbox{$p_{\mu}p^{\prime}_{\nu}$} &11.57\pm 3.12&12.55\pm3.51&1.13\pm0.34 \\
  \hline
  \mbox{$g_{\mu\nu}$} &15.24\pm 4.57&10.38\pm2.91&0.034\pm0.001 \\
                       \hline \hline
\end{array}
$$
\caption{Parameters appearing in the fit function of the coupling
form factor for $D_2^*D\pi$ vertex.} \label{fitparam}
\renewcommand{\arraystretch}{1}
\addtolength{\arraycolsep}{-1.0pt}
\end{table}
\begin{table}[h]
\renewcommand{\arraystretch}{1.5}
\addtolength{\arraycolsep}{3pt}
$$
\begin{array}{|c|c|c|c|}
\hline \hline
      \mbox{structure}    & c_1 (\mbox{GeV$^{-1}$})& c_2 (\mbox{GeV$^2$})&c_3 (\mbox{GeV$^{-1}$})    \\
\hline
  \mbox{$p_{\mu}p_{\nu}$} &6.43\pm1.92&13.31\pm3.98&-(0.79\pm0.24) \\
  \hline
  \mbox{$p^{\prime}_{\mu}p^{\prime}_{\nu}$} &9.79\pm 2.94&11.85\pm3.32&10.58\pm3.17 \\
  \hline
   \mbox{$p^{\prime}_{\mu}p_{\nu}$} &12.03\pm 3.61&12.73\pm3.18&0.81\pm0.24 \\
  \hline
  \mbox{$p_{\mu}p^{\prime}_{\nu}$} &12.03\pm 3.61&12.73\pm3.18&0.81\pm0.24 \\
  \hline
  \mbox{$g_{\mu\nu}$} &17.75\pm 5.32&10.12\pm2.84&0.062\pm0.002 \\
                        \hline \hline
\end{array}
$$
\caption{Parameters appearing in the fit function of the coupling
form factor for $D_{s2}^*DK$ vertex.} \label{fitparam1}
\renewcommand{\arraystretch}{1}
\addtolength{\arraycolsep}{-1.0pt}
\end{table}
\begin{table}[h]
\renewcommand{\arraystretch}{1.5}
\addtolength{\arraycolsep}{3pt}
$$
\begin{array}{|c|c||c|c|}
\hline \hline
     \mbox{structure}     & g_{D_2^*D\pi}(Q^2=-m_{\pi}^2)&  g_{D_{s2}^*DK} (Q^2=-m_{K}^2)   \\
\hline
  \mbox{$p_{\mu}p_{\nu}$} &4.63\pm1.39&5.76\pm1.84 \\
  \hline
  \mbox{$p^{\prime}_{\mu}p^{\prime}_{\nu}$} &20.69\pm6.21&20.59\pm5.15 \\
   \hline
  \mbox{$p^{\prime}_{\mu}p_{\nu}$} &12.72\pm3.56&12.85\pm3.85 \\
  \hline
  \mbox{$p_{\mu}p^{\prime}_{\nu}$} &12.72\pm3.56&12.85\pm3.85 \\
  \hline
  \mbox{$g_{\mu\nu}$} &15.30\pm3.67&18.26\pm5.48 \\
                        \hline \hline
\end{array}
$$
\caption{Value of the $g_{D_2^*D\pi[D_{s2}^*DK]}$ coupling
constant in $\mbox{GeV$^{-1}$}$ unit for different structures.} \label{couplingconstant}
\renewcommand{\arraystretch}{1}
\addtolength{\arraycolsep}{-1.0pt}
\end{table}

The final task in present work is to calculate the decay rates and branching ratios for the  strong $D_{2}^{*}(2460)^{0}\rightarrow D^+
\pi^- $ and $D_{s2}^{*}(2573)^{+}\rightarrow D^{+} K^{0}$ transitions. Using the amplitudes of these transitions  we find
\begin{eqnarray}\label{decaywidth}
\Gamma=\frac{|M( \mathbf{p^{\prime}})|^2}{40\pi
m^2_{D_2^*[D_{s2}^*]}}|\mathbf{p^{\prime}}|,
\end{eqnarray}
where
\begin{eqnarray}\label{M}
|M(\mathbf{p^{\prime}})|^2&=&g_{D_2^*D\pi[D_{s2}^*DK]}^2\Bigg[\frac{2}{3m_{D_2^*[D_{s2}^*]}^4}
\Big(m_{D_2^*[D_{s2}^*]}\sqrt{\mathbf{p^{\prime}}^2+m_D^2}\Big)^4
\nonumber \\
&-&\frac{4m_D^2} {3m_{D_2^*[D_{s2}^*]}^2}
\Big(m_{D_2^*[D_{s2}^*]}\sqrt{\mathbf{p^{\prime}}^2+m_D^2}\Big)^2+\frac{2m_D^4}{3}\Bigg],
\end{eqnarray}
and
\begin{eqnarray}\label{pp}
|\mathbf{p^{\prime}}|=\frac{1}{2m_{D_2^*[D_{s2}^*]}}\sqrt{m_{D_2^*[D_{s2}^*]}^4+m_D^4
+m_{\pi}^4-2m_{D_2^*[D_{s2}^*]}^2m_{\pi[K]}^2-2m_D^2m_{\pi[K]}^2-2m_{D_2^*[D_{s2}^*]}^2m_D^2}.
\end{eqnarray}
The numerical values of the decay rates for the transitions under consideration are depicted in Tables~\ref{numresult} and 6. Using the total  widths of the initial particles as
$\Gamma_{D_2^*(2460)^0}=(49.0\pm1.3)~\mbox{MeV}$,
$\Gamma_{D_{s2}^*(2573)^0}=(17\pm4)~\mbox{MeV}$ \cite{Beringer} we also find the corresponding branching ratios that
are also presented in Tables~\ref{numresult} and 6.

\begin{table}[h]
\renewcommand{\arraystretch}{1.5}
\addtolength{\arraycolsep}{3pt}
$$
\begin{array}{|c|c|c|}
\hline
 \hline
 \mbox{structure} &\Gamma(\mbox{GeV})&BR  \\
\hline
  \mbox{$p_{\mu}p_{\nu}$}  &(6.26\pm1.87)\times 10^{-4}&(1.28\pm0.36)\times 10^{-2} \\
  \hline
\mbox{$p^{\prime}_{\mu}p^{\prime}_{\nu}$} &(1.25\pm0.34)\times 10^{-2}&(2.55\pm0.74)\times 10^{-1}\\
\hline
\mbox{$p^{\prime}_{\mu}p_{\nu}$} &(4.73\pm1.42)\times 10^{-3}&(9.64\pm2.70)\times 10^{-2}\\
\hline
\mbox{$p_{\mu}p^{\prime}_{\nu}$} &(4.73\pm1.42)\times 10^{-3}&(9.64\pm2.70)\times 10^{-2}\\
\hline
\mbox{$g_{\mu\nu}$} &(5.10\pm1.48)\times 10^{-3}&(1.04\pm0.26)\times 10^{-1}\\
                        \hline \hline
\end{array}
$$
\caption{Numerical results for decay width and branching ratio of
$D_{2}^{*}(2460)^{0}\rightarrow D^+ \pi^-$ transition obtained via different structures.} \label{numresult}
\renewcommand{\arraystretch}{1}
\addtolength{\arraycolsep}{-1.0pt}
\end{table}
\begin{table}[h]
\renewcommand{\arraystretch}{1.5}
\addtolength{\arraycolsep}{3pt}
$$
\begin{array}{|c|c|c|}
\hline
 \hline
 \mbox{structure} &\Gamma(\mbox{GeV})&BR  \\
\hline
  \mbox{$p_{\mu}p_{\nu}$}  &(3.70\pm1.04)\times 10^{-4}&(2.18\pm0.59)\times 10^{-2} \\
  \hline
\mbox{$p^{\prime}_{\mu}p^{\prime}_{\nu}$} &(4.73\pm1.42)\times 10^{-3}&(2.78\pm0.69)\times 10^{-1}\\
\hline
\mbox{$p^{\prime}_{\mu}p_{\nu}$} &(1.84\pm0.48)\times 10^{-3}&(1.08\pm0.27)\times 10^{-1}\\
\hline
\mbox{$p_{\mu}p^{\prime}_{\nu}$} &(1.84\pm0.48)\times 10^{-3}&(1.08\pm0.27)\times 10^{-1}\\
\hline
\mbox{$g_{\mu\nu}$} &(3.72\pm0.97)\times 10^{-3}&(2.19\pm0.63)\times 10^{-1}\\
                        \hline \hline
\end{array}
$$
\caption{Numerical results for decay width and branching ratio of
$D_{s2}^{*}(2573)^{+}\rightarrow D^+ K^0$ transition obtained via different structures.} \label{numresult1}
\renewcommand{\arraystretch}{1}
\addtolength{\arraycolsep}{-1.0pt}
\end{table}
\begin{table}[h]
\renewcommand{\arraystretch}{1.5}
\addtolength{\arraycolsep}{3pt}
$$
\begin{array}{|c|c|c|}
\hline
 \hline
 \mbox{structure} &\Gamma(\mbox{GeV})&BR  \\
\hline
  \mbox{$p_{\mu}p_{\nu}$}  &(3.84\pm1.15)\times 10^{-4}&(7.83\pm2.03)\times 10^{-3} \\
  \hline
\mbox{$p^{\prime}_{\mu}p^{\prime}_{\nu}$} &(7.67\pm2.15)\times 10^{-3}&(1.56\pm0.44)\times 10^{-1}\\
\hline
\mbox{$p^{\prime}_{\mu}p_{\nu}$} &(2.90\pm0.87)\times 10^{-3}&(5.91\pm1.65)\times 10^{-2}\\
\hline
\mbox{$p_{\mu}p^{\prime}_{\nu}$} &(2.90\pm0.87)\times 10^{-3}&(5.91\pm1.65)\times 10^{-2}\\
\hline
\mbox{$g_{\mu\nu}$} &(3.12\pm0.75)\times 10^{-3}&(6.38\pm1.72)\times 10^{-2}\\
                        \hline \hline
\end{array}
$$
\caption{Numerical results for decay width and branching ratio of
$D_2^*(2460)^0\rightarrow D^*(2010)^+\pi^-$ transition obtained via different structures.} \label{numresult2}
\renewcommand{\arraystretch}{1}
\addtolength{\arraycolsep}{-1.0pt}
\end{table}

%
Using the following experimental ratio in $\pi$ channel \cite{Beringer,BABAR}:
\begin{eqnarray}\label{ratio}
\frac{\Gamma[D_2^*(2460)^0\rightarrow
D^+\pi^-]}{\Gamma[D_2^*(2460)^0\rightarrow
D^+\pi^-]+\Gamma[D_2^*(2460)^0\rightarrow
D^*(2010)^+\pi^-]}=0.62\pm0.03\pm0.02,
\end{eqnarray}
we  also get the values of the decay rate and branching ratio for $D_2^*(2460)^0\rightarrow
D^*(2010)^+\pi^-$ channel for different structures as presented in Table 7.

Considering  the fact that the dominant decay modes of $D_2^*(2460)$ are  $D_2^*(2460)\rightarrow
D\pi$ and  $D_2^*(2460)\rightarrow
D^{*}\pi$, from  the values presented in Tables 5 and 7, we see that all structures give the results for the total decay width of the  $D_2^*(2460)$ tensor meson compatible
with the experimental data \cite{Beringer} except for the  structure $p_{\mu}p_{\nu}$ which gives result roughly one order of magnitude smaller than the experimental values. 


To sum up,  we calculated the strong coupling form factors $g_{D_2^*D\pi}(q^2)$ and $g_{D_{s2}^*DK}(q^2)$ in the framework of QCD sum rules. Using the obtained working regions for the auxiliary parameters
entered the sum rules of the strong form factors, we found the behavior of those form factors in terms of $Q^2$. Using $Q^2=-m_{\pi[K]}^2$, we also found the values of the strong coupling constants
$g_{D_2^*D\pi}$  and $g_{D_{s2}^*DK}$ which have then been  used to calculate the decay widths and branching ratios of the strong $D_{2}^{*}(2460)^{0}\rightarrow D^+
\pi^- $, $D_2^*(2460)^0\rightarrow
D^*(2010)^+\pi^-$ and $D_{s2}^{*}(2573)^{+}\rightarrow D^{+} K^{0}$ transitions. Our results can be used in analyses of the future experimental data especially at $K$ channel.



\section{Acknowledgment}
This work has been supported in part by the Scientific and Technological
Research Council of Turkey (TUBITAK) under the research project 114F018.


\begin{thebibliography}{99}

\bibitem{Albrecht} Albrecht H, et al. (ARGUS Collaboration), Phys. Rev. Lett. 56, 549 (1986).
\bibitem{Albrecht1} H. Albrecht et al. (ARGUS Collaboration), Phys. Lett. B 232, 398 (1989);
H. Albrecht et al., (ARGUS Collaboration) Phys. Lett. B231,208
(1989); H. Albrecht et al., (ARGUS Collaboration) Phys. Lett. B221,
422 (1989);H. Albrecht et al.,(ARGUS Collaboration) Phys. Lett.
B230,162 (1989);H. Albrecht et al., (ARGUS Collaboration) Phys.
Lett. B297, 425 (1992).
\bibitem{Anjos} J. C. Anjos et al. (E691 Collaboration), Phys. Rev. Lett.62, 1717 (1989).
\bibitem{Frabetti} P. L. Frabetti et al. (E687 Collaboration),
Phys. Rev. Lett. 72, 324 (1994).
\bibitem{Avery} P. Avery et al. (CLEO Collaboration), Phys. Rev. D 41, 774 (1990);
 P. Avery et al. (CLEO Collaboration), Phys. Lett. B 331, 236 (1994).
\bibitem{Alexander} J.P. Alexander et al,(CLEO Collaboration) Phys. Lett. B 303, 377 (1993).
\bibitem{Kubota} Y. Kubota et al. (CLEO Collaboration), Phys. Rev. Lett. 72, 1972 (1994).
\bibitem{Bergfeld} T. Bergfeld et al. (CLEO Collaboration), Phys.Lett. B 340, 194 (1994).
\bibitem{Link} J. Link et al. (FOCUS Collaboration), Phys. Lett. B 586, 11 (2004).
\bibitem{Abe} K. Abe et al. (BELLE Collaboration), Phys. Rev. D 69, 112002 (2004).
\bibitem{Abazov} V. M. Abazov et al. (D0 Collaboration), Phys. Rev. Lett. 95, 161602 (2005).
\bibitem{Aaij} R. Aaij et al. (LHCb Collaboration), Phys. Lett. B 698, 14 (2011).
\bibitem{Godfrey} S. Godfrey, Phys. Rev. D 72, 054029 (2005).
\bibitem{Azizi} H. Sundu, K. Azizi, Eur. Phys. J. A48, 81 (2012).
\bibitem{Azizi1} K. Azizi, H. Sundu, J. Y. S\"{u}ng\"{u}, N. Yinelek,
 Phys. Rev. D 88,  036005 (2013); Erratum-ibid. D 88, 099901 (2013).
\bibitem{Azizi2} K. Azizi, H. Sundu, A. Y. T\"{u}rkan, E. Veli Veliev, J. Phys. G: Nucl. Part. Phys. 41, 035003 (2014).



\bibitem{Fazio} F. De Fazio, arXiv:1108.6270 [hep-ph].
\bibitem{Molina} R. Molina et. al.,
AIP Conf. Proc. 1322, 430 (2010); A. Faessler et. al., Phys. Rev. D
76, 114008 (2007).
\bibitem{Aubert1} B. Aubert et al. (BaBar
Collaboration), Phys. Rev. Lett. 103, 051803 (2009); Phys. Rev. D
80, 092003 (2009).
\bibitem{Liventsev} D. Liventsev et al. (Belle
Collaboration), Phys. Rev. D 77, 091503 (2008).
\bibitem{Colangelo} P. Colangelo et al., Phys. Rev. D 86, 054024 (2012).
\bibitem{Aubert2} B. Aubert et al., Phys. Rev. Lett. 90, 242001 (2003); D. Besson et al., Phys.
Rev. D 68, 032002 (2003) [Erratum-ibid. D 75 (2007) 119908].
\bibitem{Aubert3} B. Aubert et al., Phys. Rev. Lett. 97, 222001 (2006);
J. Brodzicka et al., Phys. Rev. Lett. 100, 092001 (2008).

\bibitem{Shifman} M. A.
Shifman, A. I. Vainshtein, V. I. Zakharov, Nucl. Phys. B147, 385
(1979); Nucl. Phys. B147, 448 (1979).
\bibitem{H.Y.Cheng} H.-Y. Cheng and K.-C. Yang, Phys. Rev. D83, 034001 (2011).



\bibitem{Reinders}L. J. Reinders, H. Rubinstein and S. Yazaki, Phys. Rept. 127, 1
(1985).


\bibitem{Beringer} Beringer et al. (Particle Data Group), Phys. Rev. D86, 010001
(2012).

\bibitem{belyaev}  V. M. Belyaev and B. L. Ioffe, Sov. Phys. JETP 57, 716 (1982); Phys. Lett. B 287, 176  (1992).
\bibitem{Falk} A. F. Falk, M. Luke, Phys. Lett. B 292, 119 (1992).
\bibitem{BABAR} The BABAR Collaboration: B. Aubert, et al, Phys.
Rev. Lett. 103, 051803 (2009).














\end{thebibliography}
\end{document}